\documentclass[11pt,twoside]{article}
\usepackage{verbatim,epsfig}
\leftline{\copyright~ 1998 International Press}
\leftline{Adv. Theor. Math. Phys. {\bf 2} (1998) {\ 1373--1404 } }

\def\npb#1#2#3{{\it Nucl.\ Phys.} {\bf B#1} (19#2) #3}
\def\plb#1#2#3{{\it Phys.\ Lett.} {\bf B#1} (19#2) #3}
\def\prl#1#2#3{{\it Phys.\ Rev.\ Lett.} {\bf #1} (19#2) #3}
\def\prd#1#2#3{{\it Phys.\ Rev.} {\bf D#1} (19#2) #3}

\def\cqg#1#2#3{{\it Class.\ Q. Grav.} {\bf #1} (19#2) #3}

\def\jhep#1#2#3{{\it J. High Energy Phys.} {\bf #1} (19#2) #3}
\def\atmp#1#2#3{{\it Adv.\ Theor.\ Math.\ Phys.} {\bf #1} (19#2) #3}
\def\endli{\hfill\break}
\def\frac#1#2{{#1 \over #2}}

\def\semi{\subset\kern-1em\times\;}
\def\bar#1{\overline{#1}}

\def\tilde#1{\widetilde#1}
\def\CA{{\cal A}}                   
                   \def\CG{{\cal G}}

                   \def\CS{{\cal S}}
\def\CU{{\cal U}}                   \def\CV{{\cal V}}

\def\C{{\bf C}}                     
                     \def\R{{\bf R}}
\def\S{{\bf S}}                     
\def\Z{{\bf Z}}                     \def\B{{\bf B}}
\def\I{{\rm I}}
\def\K{{\rm K}}
\def\KO{{\rm KO}}
\def\KR{{\rm KR}}

\def\NS{{\rm NS}}
\def\OSp{OSp\,}
\def\Sp{Sp\,}

\begin{document}
\begin{center}
\vspace{0.2in}
{\huge \bf Type IIA D-Branes, K-Theory \\[5mm] and Matrix Theory}

{\bf Petr Ho\v rava $^{a}$}
\vspace{0.2in}

$^{a}$ California Institute of Technology, \\
Pasadena, \\
CA 91125, USA\\
{\tt horava@theory.caltech.edu}
\vspace{0.2in}
\renewcommand{\thefootnote}{}
\footnotetext{\small e-print archive: {\texttt http://xxx.lanl.gov/abs/hep-th/9812135}}
\renewcommand{\thefootnote}{\arabic{footnote}}
\end{center}
\begin{abstract}

We show that all supersymmetric Type IIA D-branes can be constructed as bound 
states of a certain number of unstable non-supersym-metric Type IIA D9-branes.  
This string-theoretical construction demonstrates that D-brane charges in 
Type IIA theory on spacetime manifold $X$ are classified by the higher 
K-theory group $\K^{-1}(X)$, as suggested recently by Witten.  In particular, 
the system of $N$ D0-branes can be obtained, for any $N$, in terms of sixteen 
Type IIA D9-branes.  This suggests that the dynamics of Matrix theory is 
contained in the physics of magnetic vortices on the worldvolume of sixteen 
unstable D9-branes, described at low energies by a $U(16)$ gauge theory.  
\end{abstract}
\newpage
\pagenumbering{arabic}
\setcounter{page}{1374}

\pagestyle{myheadings}
\markboth{\it TYPE IIA D-BRANES, K-THEORY, AND MATRIX THEORY}{\it PETR HO\v RAVA}
\section{Introduction}
\setcounter{equation}{0}

When we consider individual D-branes in Type IIA or Type IIB string theory on 
$\R^{10}$, we usually require that the branes preserve half of the original 
supersymmetry, and that they carry one unit of the corresponding RR charge.  
These requirements limit the D-brane spectrum to $p$-branes with all even 
values of $p$ in Type IIA theory, and odd values of $p$ in Type IIB theory. 

Once we relax these requirements, however, we can consider D$p$-branes with 
all values of $p$.  In Type IIA theory, we can consider $p$-branes with $p$ 
odd, and in particular, a spacetime-filling 9-brane.  All these states are 
non-supersymmetric unstable excitations in the corresponding supersymmetric 
string theory.  Indeed, there is always a tachyon in the spectrum of the open 
string connecting one such Type IIA $(2p-1)$-brane to itself.  Thus, such 
D-brane configurations (and their counterparts on spacetimes of non-trivial 
topology) are highly unstable, and one expects that they should rapidly decay 
to the supersymmetric vacuum, by a process that involves tachyon condensation 
on the worldvolume.  This is of course in perfect agreement with the field 
content of the corresponding low-energy supergravity in spacetime -- there 
are no RR fields that could couple to any conserved charges carried by such 
non-supersymmetric branes.  

This does not seem to leave much room for surprises, but in fact, the full 
story is much more interesting.  Configurations of 
unstable D-branes can sometimes carry lower-dimensional D-brane charges, and 
therefore, when the tachyon rolls down to the minimum of its potential and 
the state decays, it can leave behind a supersymmetric state that differs 
from the vacuum by a lower-dimensional D-brane charge -- in other words, the 
state decays into a supersymmetric D-brane configuration.  Typically, one can 
then represent the supersymmetric D-brane state as a bound state of the 
original system of unstable D-branes.  

This setup generalizes a special case studied in \cite{sen,ewk}, where one 
starts with an unstable configuration of an equal number of stable $p$-branes 
and stable anti-$p$-branes (or $\bar p$-branes for short), and finds 
lower-dimensional stable D-brane states as bound states in this system.  
The $p$-brane $\bar p$-brane system is unstable, and 
the instability manifests itself by the presence of a tachyon in the spectrum 
of the $p$-$\bar p$ open strings at brane separations shorter than the string 
scale \cite{tbls}-\cite{srednicki}.  In Type IIB or Type I string theory, one can 
use this construction to represent any stable D-brane state as a bound state 
of a certain number of 9-brane $\bar 9$-brane pairs \cite{sen,ewk}.

In some cases, the unstable non-supersymmetric state decays into a stable 
state that is not supersymmetric, but is protected from further decay by 
charge conservation.  
One typical example of such states is the $SO(32)$ spinor of Type I theory 
\cite{sen}, which is non-supersymmetric but stable, since it is the 
lowest spinor state in the theory.  Such non-supersymmetric D-branes can 
be found using a direct boundary state construction \cite{berggab} , or 
alternatively as bound states of $p$-brane $\bar p$-brane pairs \cite{sen} .  This 
remarkable bound-state construction was discovered by Sen \cite{sen} (following 
some earlier work in \cite{boundictp} ), and was further systematized by Witten 
\cite{ewk} .  It turns out that the proper setting for all conserved D-brane 
charges in general compactifications is, in fact, K-theory \cite{rmgm,ewk}.  
(Some early indications of possible connections between D-branes and K-theory 
can be found in \cite{phdthesis}-\cite{cheungyin}.)  

In this paper, we generalize the construction that uses unstable 
configurations of pairs of stable branes, and consider bound states in 
general unstable non-supersymmetric configurations of D-branes such as the 
unstable 9-brane discussed above, where the individual D-branes are no longer 
required to be stable.  

Our motivation for this generalization will be clear from the following 
example.  We will be interested in stable D-branes in Type IIA string theory, 
for simplicity in 
$\R^{10}$.  In Type IIB theory, we can in principle construct any stable 
D-brane state as a bound state of a certain number of 9-brane $\bar 9$-brane 
pairs wrapping the whole spacetime.  It is certainly desirable to have, on  
the Type IIA side, a similar construction that would enable us to study 
stable D-branes as bound states in unstable configurations of branes of 
maximal dimension.  However, there are no stable D9-branes in Type IIA string 
theory!  While we can indeed represent any stable D-brane of Type IIA as a 
bound state of an 8-brane $\bar 8$-brane system, this construction requires a 
preferred choice of a submanifold of codimension one in space-time, 
representing the worldvolume of the 8-brane $\bar 8$-brane system.    
Therefore, it breaks some of the spacetime symmetries that we want to keep 
manifest in the theory, and limits the kinematics of branes that can be 
studied this way.  One of the main points of this paper is to present 
a string-theoretical construction that keeps all spacetime symmetries 
manifest.  This construction enables us to consider any stable D-brane 
of Type IIA theory as a bound state of a system of the unstable 9-branes 
discussed above.  In fact, this construction turns out to be intimately 
related to the statement that D-brane charges in Type IIA string theory are 
classified by the higher K-theory group $\K^{-1}(X)$ of the spacetime manifold 
$X$, suggested recently by Witten in \cite{ewk} .  

This paper is organized as follows.  

In section~2 we briefly review the relation of Type IIB D-brane charges and 
bound-state constructions to K-theory, and preview the Type IIA case.  

In section~3.1 we introduce the unstable 9-brane of Type IIA string theory.  
General 9-brane configurations wrapping spacetime manifold $X$ are studied 
in section~3.2.  We argue that inequivalent configurations of 9-branes -- 
modulo 9-branes that can be created from or annihilated to the vacuum -- 
are classified by the higher K-theory group $\K^{-1}(X)$.  In section~3.3 we 
show that any given stable D-brane configuration of Type IIA string theory can 
be represented as a bound state of a certain number of unstable Type IIA 
9-branes.  In the worldvolume of the 9-branes, the bound state appears as a 
stable vortex in the tachyon field, accompanied by a non-trivial gauge field 
carrying a generalized magnetic charge.  In the particular case of bound 
states in codimension three, this precisely corresponds to the 
't~Hooft-Polyakov magnetic monopole.  We generalize our discussion to the 
case of Type $\I'$ theory in section~3.4, and argue that Type $\I'$ D-brane 
charges are similarly classified by the Real K-theory group $\KR^{-1}(X)$.  

In section~4 we focus on possible implications of our construction to Matrix 
theory \cite{bfss} .  We use our 9-brane bound-state construction to study a 
general system of $N$ D0-branes in Type IIA theory.  {}First we show that a 
D0-brane can be constructed as a bound state of sixteen unstable D9-branes.  
The low-energy worldvolume theory on the 9-branes is a certain 
non-supersymmetric $U(16)$ gauge theory, with a tachyon in the adjoint 
representation of $U(16)$.  In this worldvolume theory, the D0-brane is 
represented as a topologically stable vortex-monopole configuration of the 
tachyon and the gauge field.  Multiple D0-brane configurations are in general 
also described by sixteen 9-branes, and appear as multi-vortex configurations 
in the $U(16)$ gauge theory on the spacetime-filling worldvolume.  This 
construction thus leads to the intriguing possibility that the dynamics of 
Matrix theory -- as described by a particular limit of the system of $N$ 
D0-branes of Type IIA string theory -- can be contained in the dynamics of 
vortices on the worldvolume of a fixed system of sixteen D9-branes, with the 
individual D0-branes represented by vortices in the worldvolume field theory.  
This construction exhibits some striking similarities with the holographic 
field theory of \cite{hft} .  

While this paper was being finished, another paper appeared \cite{sennew} whose 
section~6 partially overlaps with some parts of our section~3.3. 

\section{K-Theory and Type IIA D-Branes}
\setcounter{equation}{0}

In this section we first review some highlights of \cite{ewk} (mostly in the 
context of Type IIB theory), which will give us the opportunity to present 
some background on K-theory \cite{karoubi}-\cite{atiyah} that will be useful later 
in the paper.  In section~2.2 we set the stage for our further discussion of 
D-brane charges and bound states in Type IIA theory. 

\subsection{Type IIB on $X$ and $\K(X)$}

Consider supersymmetric $p$-branes and $\bar p$-branes of Type IIB (or Type 
IIA) string theory, wrapped on a spacetime manifold $X$ of dimension $p$.  
The lowest states of the open strings connecting $N$ such $p$-branes give 
rise to a worldvolume $U(N)$ gauge field on a Chan-Paton bundle $E$ of complex 
dimension $N$.  Similarly, $N'$ $\bar p$-branes will carry a Chan-Paton bundle 
$E'$ of dimension $N'$ and a $U(N')$ gauge field.  The open string connecting 
a $p$-brane to a $\bar p$-brane will have the opposite GSO projection, and 
its lowest mode will be a tachyon field $T$ in $(N,\bar{N'})$ of the gauge 
group $U(N)\times U(N')$.  

The D-brane charge of the configuration will be preserved in processes where 
$p$-brane $\bar p$-brane pairs are created from or annihilated to the vacuum.  
Configurations that can be created or annihilated correspond to $N$ $p$-branes 
with bundle $F$ and $N$ $\bar p$-branes with a bundle $F'$ that is 
topologically equivalent to $F$.  Thus, invariant D-brane charges correspond 
to equivalence classes of pairs of bundles $(E,E')$, where two pairs 
$(E_1,E_1')$ and $(E_2,E_2')$ are equivalent if $(E_1\oplus F,E_1'\oplus F)$ 
is isomorphic to $(E_2\oplus G,E_2'\oplus G)$ for some $F$ and $G$.  (If 
$F$ corresponds to brane antibrane pairs being created from the vacuum, $G$ 
corresponds to pairs annihilated to the vacuum.)  The set $\K(X)$ of such 
equivalence classes of pairs of bundles on $X$ forms a group, called the 
K-theory group of $X$.  The image of $(E,0)$ in $\K(X)$ is sometimes denoted 
by $[E]$.  Each element in $\K(X)$ can be written as $[E]-[E']$ for some 
bundles $E$ and $E'$.  

Consider configurations of 9-branes and $\bar 9$-branes in Type IIB 
string theory, wrapping a spacetime manifold $X$.  Tadpole cancellation 
requires an equal number of branes and antibranes.  In K-theory, one can show 
-- for $X$ a connected compact manifold -- that $\K(X)$ canonically factorizes 
as $\tilde\K(X)\oplus\Z$, where the integer in $\Z$ is simply $N-N'$, the 
difference between the number of $p$-branes and the number of $\bar p$-branes. 
{}For a non-compact manifold $Y$, one defines $\K(Y)=\tilde\K(\tilde Y)$, 
where $\tilde Y$ is a compactification of $Y$ by adding a point at infinity.  
Thus, on a general spacetime manifold $X$, D-brane charges of tadpole 
cancelling Type IIB 9-brane $\bar 9$-brane configurations are classified by 
the reduced K-theory group $\tilde\K(X)$.  

Each class in $\tilde\K(X)$ is represented by a system with an equal number 
of 9-branes and $\bar 9$-branes wrapping $X$, with the class in $\K(X)$ given 
by the difference of the Chan-Paton bundles on 9-branes and $\bar 9$-branes, 
$[E]-[E']$.  Open strings ending on all possible pairs of these branes will 
give rise to a $U(N)\times U(N)$ gauge field, and the tachyon field $T$ in 
the $(N,\bar N)$ representation of the gauge group.  Together, these bosonic 
fields form an object 
\begin{equation}\label{eesuperconn}
\pmatrix{A&T\cr\bar T&A'\cr},
\end{equation}
\noindent known in the mathematical literature as a ``superconnection'' on $X$ 
\cite{quillen} .  

The tachyon has the tendency to roll down to the minimum of its potential 
located at some $T=T_0$.  We do not know the exact form of the tachyon 
potential, but it was argued in \cite{ewk} that at the minimum of the potential, 
all eigenvalues of $T_0$ are equal, and therefore $T_0$ breaks the gauge 
symmetry from $U(N)\times U(N)$ to the diagonal $U(N)$.  

Any lower-dimensional stable D-brane of Type IIB theory, wrapping a 
submanifold $Y$ in spacetime, can be constructed as a bound state in a system 
of $N$ 9-branes and $N$ $\bar 9$-branes, which locally near $Y$ looks like a 
topologically stable vortex of the tachyon field.
\footnote{For the purposes of this paper, it will be sufficient to consider 
only branes stretching along submanifolds $\R^m$ of the flat spacetime 
$\R^{10}$.  The general case of Type IIB D-branes wrapping general 
submanifolds $Y$ in general spacetime $X$ is discussed in \cite{ewk} .}
This can be seen as follows.  Stable values of $T_0$ correspond to the vacuum 
manifold 
\begin{equation}\label{eevack}
\CV_0(N)=(U(N)\times U(N))/U(N),
\end{equation}
\noindent which is topologically equivalent to $U(N)$.  Thus, the tachyon will support 
stable defects in codimension $2k$, classified by the non-zero homotopy 
groups of the vacuum manifold, $\pi_{2k-1}(\CV_0(N))=\Z$ (for stable values 
of $N$).  In order for these defects to carry finite energy, the vortex of 
winding number $n$ in the tachyon field must be accompanied by a non-trivial 
gauge field configuration carrying $n$ units of the corresponding topological 
charge.  In the simplest case of one $p$-brane $\bar p$-brane pair, a vortex 
of codimension two with vorticity $n$ carries $n$ units of the magnetic flux 
in one of the $U(1)$ groups. 

The core of the vortex lies along a submanifold $Y$ of codimension $2k$ in 
$X$.  Outside a small core around the submanifold $Y$ in $X$, the tachyon 
condenses to its vacuum value.  It is believed \cite{sen,boundictp} that 
the negative energy density corresponding to the vacuum condensate of the 
tachyon field is equal in magnitude to the positive energy density due to 
the non-zero tension of the $p$-brane $\bar p$-brane system wrapping $X$.  
Consequently, the total energy density away from the core of the bound state 
rapidly approaches zero, and the configuration is very close to the 
supersymmetric vacuum.  Thus, the process of tachyon condensation leaves 
behind an object wrapped on $Y$ that carries the charge of a supersymmetric 
D-brane wrapping $Y$ \cite{sen} . 

In terms of K-theory, this construction corresponds to the embedding of
a non-trivial class in $\K(Y)$ -- describing a D-brane wrapping $Y$ -- 
into $\K(X)$, where it corresponds to the bound state of a 9-brane 
$\bar 9$-brane configuration wrapping $X$.  This embedding is realized by a 
classic K-theory construction \cite{abs} , which selects -- 
for a $Y$ of codimension $2k$ in $X$ -- a preferred value of the number of 
9-brane $\bar 9$-brane pairs $N=2^{k-1}$, and which also leads to a 
particularly natural representative of the tachyon vortex configuration.  
The tachyon vortex that (at least locally in $X$) represents the D-brane 
wrapping a manifold $Y$ of codimension $2k$ in $X$ is constructed as 
follows.  The group of rotations $SO(2k)$ in the dimensions normal to $Y$ in 
$X$ has two inequivalent spinor representations, which give rise to two spinor 
bundles $\CS_+$ and $\CS_-$ on $Y$.  These bundles can be extended to a 
neighborhood of $Y$ in $X$, and therefore they define (modulo possible global 
obstructions that can typically be eliminated by pair creating extra 9-branes 
and $\bar 9$-branes) a K-theory class $[\CS_+]-[\CS_-]$.  In the construction 
of the bound state, we identify $\CS_+$ with the Chan-Paton bundle carried by 
9-branes, and $\CS_-$ with the bundle supported by $\bar 9$-branes.  The gauge 
symmetry on the 9-brane worldvolume is $U(2^{k-1})\times U(2^{k-1})$.  The 
tachyon is a map from $\CS_+$ to $\CS_-$, and its vortex configuration of 
vorticity one around $Y$ can be written in a very simple form using $\Gamma$ 
matrices of $SO(2k)$ (which represent maps from $\CS_+\oplus\CS_-$ to itself), 
as 
\begin{equation}\label{eetachk}
T=\Gamma_m x^m,
\end{equation}
\noindent where $x^m$, $m=1,\ldots 2k$ are coordinates in the directions transverse to 
$Y$ in a small neigborhood of $Y$ in spacetime.  We have omitted a 
multiplicative convergence factor in \ref{eetachk} , which approaches one in the 
small vicinity of $Y$ and goes to 
zero as we approach infinity, ensuring that far away from the core of the 
vortex, $T$ takes values in the vacuum manifold \ref{eevack} , with $N=2^{k-1}$.  
\footnote{Our discussion has been local in $X$; when global topology is taken into 
account, one sometimes has to ``stabilize'' (in the K-theory sense) the 
configuration of 9-branes and $\bar 9$-branes by pair creating extra 9-branes 
and $\bar 9$-branes, thus leading to a configuration of 9-brane $\bar 9$-brane 
pairs described by $(\CS_+\oplus H,\CS_-\oplus H)$; for more details, see 
\cite{ewk} .}

As an example, let us consider the case of codimension two.  The gauge 
symmetry is $U(1)\times U(1)$, and the tachyon vortex of vorticity one can be 
written as 
\begin{equation}\label{eetachcodimtwo}
T=\sigma_ix^i=\pmatrix{0&x^1+ix^2\cr x^1-ix^2&0\cr},
\end{equation}
\noindent where we have used a particular convenient representation of the two 
$\Gamma$-matrices $\sigma_{1,2}$ in two transverse dimensions $x^{1,2}$.  

This bound-state construction defines a map $\K(Y)\rightarrow\K(X)$ for 
any submanifold $Y$ of $X$ that admits a Spin${}_c$ structure \cite{ewk} .  Some 
other details of this construction, together with more details about its 
relation to the Thom isomorphism, the Gysin map, and the Atiyah-Bott-Shapiro 
construction in K-theory, can be found in \cite{ewk} .  For some general K-theory 
background, see \cite{karoubi}-\cite{atiyah}.

\subsection{Type IIA on $X$ and $\K^{-1}(X)$}

It has been suggested in \cite{ewk} that D-brane charges in Type IIA theory should 
be similarly classified by a certain higher K-theory group $\K^{-1}(X)$.   
This conjecture is supported by the following argument.  Consider the reduced 
K-theory groups of spheres, $\tilde\K(\S^n)$.  These groups classify possible 
$(9-n)$-branes in Type IIB theory on $\R^{10}$ \cite{ewk} .  Using Bott periodicity, 
one can show that $\tilde\K(\S^{2n})=\Z$ and $\tilde\K(\S^{2n+1})=0$.  The 
higher K-theory group $\K^{-1}(X)$ will be defined precisely below, but now we 
only invoke the fact that 
\begin{equation}\label{eekonekspheres}
\K^{-1}(\S^n)=\tilde\K(\S^{n+1}).
\end{equation}
\noindent Hence, $\K^{-1}(\S^{2n+1})=\Z$ and $\K^{-1}(\S^{2n})=0$.  This is in accord 
with the fact that Type IIB theory contains supersymmetric $p$-branes for $p$ 
odd, while Type IIA theory has $p$-branes with $p$ even.  

Thus, the higher K-theory group $\K^{-1}(X)$ of spacetime $X$ is a natural 
candidate for the K-theory group that classifies D-brane charges in Type IIA 
theory.  
{}For a manifold $X$ of dimension $d$, the higher K-theory group $\K^{-1}(X)$ 
is usually defined using the ordinary K-theory group of a $d+1$ dimensional 
extension $X'$ of $X$.  If $X$ is a spacetime manifold of string theory, $X'$ 
will be eleven-dimensional, and we may suspect a connection to M-theory.  The 
definition of $\K^{-1}(X)$ that is most suggestive of M-theory sets 
$X'=X\times\S^1$, and defines $\K^{-1}(X)$ using the K-theory group 
$\tilde\K(X\times\S^1)$.  More precisely, $\K^{-1}(X)$ is defined as the 
subgroup in $\tilde\K(X\times\S^1)$ that maps to the trivial class in 
$\tilde\K(X)$, by the map induced from the embedding of $X$ as 
$X\times{\rm point}$ in $X\times\S^1$.%
\footnote{The purpose of this extra condition is to eliminate elements in 
$\tilde\K(X\times\S^1)$ that correspond to $\tilde\K(X)$.  In the string 
theory language, since $\K^{-1}(X\times\S^1)\cong\K^{-1}(X)\oplus\tilde\K(X)$, 
the restriction of $\tilde\K(X\times\S^1)$ to its subgroup $\K^{-1}(X)$ 
eliminates charges that would correspond in Type IIA string theory to stable 
$p$-branes with $p$ odd.}
This definition of $\K^{-1}(X)$ that uses $X'=X\times\S^1$ is somewhat 
awkward, and we can define $\K^{-1}(X)$ more directly by choosing a slightly 
different $X'$ as follows.  Consider first the product of $X$ with a unit 
interval, $X\times I$, and define the so-called ``suspension'' $S'(X)$ of $X$ 
by identifying all points in each boundary component of $X\times I$.  Thus, 
for example, the suspension of the $m$-sphere $\S^m$ is the 
$(m+1)$-dimensional sphere, $S'(\S^m)=\S^{m+1}$.  One can define $\K^{-1}(X)$ 
by starting with $X'=S'(X)$, and setting 
\begin{equation}\label{definkone}
\K^{-1}(X)=\tilde\K(S'(X)).
\end{equation}
\noindent In the particular case of $X=\S^m$, we obtain \ref{eekonekspheres} .  
 
In Type IIB theory, the fact that $\tilde\K(X)$ classifies D-brane charges  
leads to the construction of all possible D-branes as bound states of 
spacetime-filling 9-brane $\bar 9$-brane pairs, reviewed briefly in the 
previous section.  When we try to extend this construction of $\K(X)$ from 
Type IIB theory to an analogous construction of the $\K^{-1}(X)$ groups on 
the Type IIA side, we encounter the following difficulty.  As we have seen, 
$\K^{-1}(X)$ is defined as the usual K-theory group of an extended manifold 
$S'(X)$.  In string theory, $X$ is a spacetime manifold of dimension ten, and 
therefore its extension $S'(X)$ used in the definition of $\K^{-1}(X)$ is 
eleven-dimensional.  This indeed suggests a possible relation to M-theory (as 
pointed out in \cite{ewk}), but since we do not have a hierarchy of D-branes 
in M-theory (and in particular there are no 10-branes), it is unclear how to 
interpret the $\K^{-1}(X)$ group that should classify D-brane charges of 
Type IIA string theory but is defined using an eleven-dimensional extension 
of $X$.  

In the next section we will present a string theory construction (as opposed 
to an M-theory construction) of all supersymmetric D-branes of Type IIA string 
theory, similar in spirit to that of \cite{sen,ewk}.  The unstable 9-branes 
of Type IIA theory, introduced in section~1, will provide the crucial 
ingredient for our construction.  
We will see that this stringy construction is a translation of one possible 
definition of $\K^{-1}(X)$ in K-theory.  Hence, our construction proves that 
the D-brane charges of Type IIA string theory are indeed classified by 
$\K^{-1}(X)$, as suggested by Witten in \cite{ewk} .  The argument can be easily 
generalized to see that D-branes in Type $\I'$ theory are similarly classified 
by $\KR^{-1}(X)$.  
\vfill
\section{Type IIA D-Branes as Bound States of Unstable 9-Branes}
\setcounter{equation}{0}
\subsection{Unstable 9-Branes in Type IIA Theory}
We have pointed out in section~1 that once we relax the condition that 
D-branes be supersymmetric and carry a RR charge, we can construct D$p$-branes 
of any $p\leq 9$, at the cost of sometimes obtaining unstable configurations.  
In particular, in Type IIA theory we can construct a 
spacetime-filling 9-brane.  Its structure can be easily understood from the 
form of its boundary state $\left|B\right\rangle$, which -- as a particular 
coherent state in the Hilbert space of the Type IIA closed string -- 
represents the boundary conditions on the closed string annihilated into the 
9-brane.  The D9-brane boundary state (we will only consider the 9-brane of 
Type IIA theory, generalizations to $(2k-1)$-branes of lower dimensions are 
obvious) is thus given by 
\begin{equation}\label{eeb}
\left|B\right\rangle=\left|B,+\right\rangle_{\NS\,\NS}-\left|B,-
\right\rangle_{\NS\,\NS}.
\end{equation}
\noindent Here $\left|B,\pm\right\rangle_{\NS\,\NS}$ represents the two possible 
implementations of the Neumann boundary conditions on all spacetime 
coordinates, \cite{polcai}-\cite{wso}. 

There is no RR component in the boundary state, as none is invariant under the 
Type IIA GSO projection in the closed string channel.  Indeed, there are two 
RR states $\left|B,\pm\right\rangle_{\rm RR}$ that implement Neumann boundary 
conditions on all coordinates.  These states transform into each other under 
the worldsheet fermion number operators $(-1)^{F_{L,R}}$, as follows (see 
e.g.\ \cite{berggab} ):
\begin{equation}\label{eetransfermi}
(-1)^{F_{L}}\left|B,\pm\right\rangle_{\rm RR}=
\left|B,\mp\right\rangle_{\rm RR},\qquad(-1)^{F_R}\left|B,\pm\right
\rangle_{\rm RR}=\left|B,\mp\right\rangle_{\rm RR}.
\end{equation}
\noindent However, the GSO projection in Type IIA theory chooses opposite chiralities 
in the left-moving and the right-moving sector, and no combination of 
$\left|B,\pm\right\rangle_{\rm RR}$ is invariant under the Type IIA GSO 
operator $(1-(-1)^{F_L})(1+(-1)^{F_R})$.%
\footnote{I am grateful to Oren Bergman for discussions on this subject.}
The absence of a RR boundary state means that no RR tadpole is associated 
with our 9-brane, and therefore, no spacetime anomalies related to RR 
tadpoles can arise.  Unlike in Type IIB theory, where tadpole cancellation 
requires an equal number of 9-branes and $\bar 9$-branes, we do not get a 
restriction on the number of 9-branes from tadpole cancellation.  These 
9-branes carry no conserved charge, and there is no distinction between a 
9-brane and an anti-9-brane.  

Since there is no RR component in the boundary state, there is no GSO 
projection in the dual, open-string channel of the toroidal amplitude 
$\left\langle B|B\right\rangle$.  Therefore, the open string connecting one 
such 9-brane to itself will contain -- in the NS sector -- both the $U(1)$ 
gauge field that a supersymmetric brane would carry, and the tachyon field $T$ 
that would, in the case of supersymmetric branes, be projected out by the GSO 
projection.  In the Ramond sector of the open string, both spacetime 
chiralities of the ground state spinor are retained, again due to the absence 
of any GSO projection.  

A more precise way of implementing this boundary-state construction of 
unstable D-branes in a way compatible with the general Type IIA GSO 
projections on higher-genus worldsheets has been proposed in a similar case 
of unstable Type IIB D0-branes by Witten in \cite{ewk} .  In this 
procedure, one introduces an extra fermion $\eta$ at each boundary component 
that corresponds to the string worldsheet ending on the 9-brane.  (Similar 
boundary fermions were introduced some time ago in a different context in 
\cite{marcsag} ).  This extra fermion is described by the Lagrangian 
$\int\eta(d\eta/dt)dt$, with $t$ a periodic coordinate along the 
worldsheet boundary component.  Quantization of this fermion gives an extra 
factor of $\sqrt 2$ in the sector with antiperiodic boundary conditions, 
and the zero mode of $\eta$ kills the contribution to the worldsheet path 
integral from the R sector with periodic boundary conditions.  In terms of 
boundary states, this indeed reproduces our boundary state 
$\left| B\right\rangle$, but now not because there would be no GSO 
projection.  Instead, there is a GSO projection, the Ramond boundary state 
$\left| B,R\right\rangle$ vanishes identically, and the total 
boundary state can be rewritten as 
\begin{equation}\label{eetotalboundeta}
\left|B\right\rangle=\frac{1}{\sqrt 2}\left(
\sqrt 2\left|B\right\rangle+\left|B,R\right\rangle\right).
\end{equation}
\noindent The extra factor of $\sqrt 2$ comes from the extra Chan-Paton factor 
associated with the extra boundary fermion $\eta$.

In the presence of Type IIA $2p$-branes carrying RR charges, there would 
be a sector of open strings connecting the $2p$-brane to the 9-brane.  As 
in the case of Type IIB D0-branes studied in \cite{ewk} , the worldsheet rules for 
calculating amplitudes in such cases also require the presence of the extra 
boundary fermion $\eta$.  

More generally, consider $N$ coincident 9-branes.  The free open-string 
spectrum of massless and tachyonic states gives rise to the following 
low-energy field content on the spacetime-filling worldvolume,
\begin{equation}\label{eespectnine}
A_\mu,\ T,\ \chi,\ {\rm and}\ \chi',
\end{equation}
\noindent where $A_\mu$ is a $U(N)$ gauge field, $T$ is a tachyon field in the adjoint 
of $U(N)$, and the two chiral fermions $\chi,\chi'$ -- also in the adjoint of 
$U(N)$ -- carry opposite spacetime chiralities.%
\footnote{In the simplest case of just one 9-brane, the gauge group is $U(1)$, 
and the tachyon is a real scalar field.  We will see below that this case is 
somewhat degenerate, and one may want to ``stabilize'' it -- in the sense 
of K-theory -- by embedding this system into a larger system with more than 
one 9-brane.}
This should be contrasted with the worldvolume field content of the Type IIB 
system of $N$ pairs of 9-branes and $\bar 9$-branes, where the bosonic sector 
\ref{eesuperconn} consists of a $U(N)\times U(N)$ gauge field and a tachyon 
in $(N,\bar N)$.  

Notice the intriguing fact that this field content \ref{eespectnine} on $N$ 
9-branes of Type IIA theory coincides with the ten-dimensional decomposition 
of a system in eleven dimensions, consisting of a $U(N)$ gauge field $\CA_M$ 
and a 32-component spinor $\Psi$ in the adjoint of $U(N)$.  In particular, 
the adjoint tachyon plays the role of an eleventh component of the $U(N)$ 
gauge field, and the ten-dimensional decomposition gives \ref{eespectnine} as 
\begin{equation}\label{eetendecomp}
\CA_M=(A_\mu,T),\qquad\Psi=(\chi,\chi').
\end{equation}
\noindent Of course, this hidden eleven-dimensional symmetry of the lowest open-string 
states is broken already at the level of free fields by the tachyon mass.  

\subsection{9-Brane Configurations and $\K^{-1}(X)$}

In analogy with our understanding of Type IIB D-branes in K-theory, we want 
to achieve two separate things: (1) classification of branes in Type 
IIA theory on general $X$, (2) construction of branes in terms of bound 
states of higher-dimensional branes.  

First, we will consider possible configurations of $N$ 9-branes in Type IIA 
string theory, up to possible creation and annihilation of 9-branes from and 
to the vacuum.  

Recalling our discussion in Section~1 of a system of such unstable 9-branes 
in Type IIA theory, we expect that the system will rapidly decay to the 
supersymmetric vacuum, whenever it does not carry lower-dimensional D-brane 
charges.  We will call such 9-brane configurations ``elementary.''  Any 
such ``elementary'' configuration of $N'$ branes wrapping $X$ will give rise 
to a $U(N')$ bundle $F$, together with a $U(N')$ gauge field on $F$ and a 
tachyon $T$ in the adjoint representation of $U(N')$.  The bound-state 
construction that we discuss below indicates that the presence or absence of 
lower D-brane charges can be measured by the tachyon condensate $T$.  

Thus, we will assume (cf.\ \cite{sen,ewk}) that a bundle $E$ with tachyon 
field $T$ can be deformed -- by processes that involve only creation and 
annihilation of ``elementary'' 9-branes -- into a bundle isomorphic to 
$E\oplus F$, with $F$ the Chan-Paton bundle of an elementary 9-brane 
configuration.  

This definition of equivalence classes of 9-branes with tachyon condensate, 
up to creation or annihilation of ``elementary'' 9-brane configurations from 
and to the vacuum, corresponds to the following construction in K-theory.  

It turns out \cite{karoubi} that in K-theory, one can define the higher K-theory 
group $\K^{-1}(X)$ without using an eleven-dimensional extension of $X$.  
Instead, one starts with pairs 
$(E,\alpha)$, where $E$ is a $U(N)$ bundle for some $N$, and $\alpha$ is 
an automorphism on $E$.  (In fact, we do not lose generality if we consider 
only trivial bundles $E$ on $X$.)  
A pair $(F,\beta)$ is called ``elementary'' if the automorphism $\beta$ can 
be continuously deformed to the identity automorphism on $F$, within 
automorphisms of $F$.  One defines an equivalence relation on pairs 
$(E,\alpha)$, as follows.  Two pairs $(E_1,\alpha_1)$ and $(E_2,\alpha_2)$ 
are equivalent if there are two {\it elementary\/} pairs $(F_1,\beta_1)$ and 
$(F_2,\beta_2)$ such that 
\begin{equation}\label{eestringdefkone}
(E_1\oplus F_1,\alpha_1\oplus\beta_1)\cong
(E_2\oplus F_2,\alpha_2\oplus\beta_2).
\end{equation}
\noindent The set of all such equivalence classes of pairs $(E,\alpha)$ on $X$ is a 
group: the inverse element to the class of $(E,\alpha)$ is the class of 
$(E,\alpha^{-1})$.  This group of classes of pairs $(E,\alpha)$ on $X$ is 
precisely $\K^{-1}(X)$ (as defined e.g.\ in \cite{karoubi} , Section~II.3).  
This ``string theory'' definition of $\K^{-1}(X)$ -- which uses bundles with 
automorphisms on the ten-dimensional spacetime $X$ -- is equivalent to the 
definition of $\K^{-1}(X)$ reminiscent of M-theory (and reviewed in 
section~2.2) which uses pairs of bundles on the eleven-dimensional extension 
$X\times\S^1$.  This rather non-trivial fact can be found e.g.\ in \cite{karoubi} , 
Theorem~II.4.8. 

In string theory, the role of the $N$-dimensional bundle $E$ is played by 
the Chan-Paton bundle carried by a system of $N$ unstable Type IIA 9-branes.  
The automorphism $\alpha$ is a little harder to see directly in the 9-brane.   
However, we will see below that in the bound-state construction of 
supersymmetric D-branes as bound states in a system of 9-branes, the role of 
$\alpha$ is played by 
\begin{equation}\label{eeroleofauto}
\CU=-e^{\pi iT},
\end{equation}
\noindent where $T$ is the adjoint $U(N)$ tachyon on the 9-brane worldvolume.  
Elementary pairs $(F,\alpha)$ correspond to elementary brane configurations 
that do not carry any lower-dimensional D-brane charge, and therefore can be 
created from and annihilated to the vacuum.  Thus, possible 9-brane 
configurations up to creation and annihilation of ``elementary'' 9-branes are 
classified by $\K^{-1}(X)$.  This, together with our explicit bound-state 
construction below, demonstrates that $\K^{-1}(X)$ indeed classifies D-brane 
charges in Type IIA theory.  

In contrast to Type IIB theory, where one is supposed to consider tadpole 
cancelling configuration of an equal number of 9-branes and $\bar 9$-branes, 
in Type IIA theory there is no such restriction on the number of 9-branes.  
This statement has a nice interpretation in K-theory.  In Type IIB theory, 
tadpole cancelling configurations of 9-brane $\bar 9$-brane pairs correspond 
to the reduced K-theory group, $\tilde K(X)$, related to the full $\K(X)$ by 
$\K(X)=\Z\oplus\tilde\K(X)$.  One can define a ``reduced'' higher K-theory 
group $\tilde\K{}^{-1}(X)$ \cite{karoubi,atiyah}, but it turns out that 
(for the class of spacetime manifolds that one encounters in string theory) 
$\tilde\K^{-1}(X)$ is always equal to $\K^{-1}(X)$.  

\subsection{Type IIA D-Branes as Bound States of 9-Branes}

So far, we have suggested a classification of all configurations of 9-branes 
up to creation or annihilation of ``elementary'' 9-branes that do not carry 
any lower D-brane charge.  Here we present a construction that allows one to 
embed any lower-dimensional branes into a system of 9-branes in Type IIA 
theory: thus, just as in Type IIB theory \cite{ewk} , whatever can be done with 
stable lower-dimensional branes can be done with unstable 9-branes of Type 
IIA theory.  

Even though we will mostly focus on bound states of unstable 9-branes, 
one could also start with any lower-dimensional unstable $(2k-1)$-branes, 
and construct stable $2p$-branes for $p\leq k-1$ as their bound states.  
In turn, each such lower-dimensional unstable $(2k-1)$-brane can be viewed as 
an unstable bound state of a $2k$-brane $\bar{2k}$-brane pair, and we obtain 
a whole hierarchy of bound-state constructions, corresponding to a 
hierarchy of K-theory isomorphisms.  However, the only truly interesting case 
is that of 9-branes, for the following reason.  Using the techniques of \cite{ewk} , 
each individual lower-dimensional stable D$p$-brane of Type IIA theory can 
already be constructed as a bound state of a certain number of 8-brane 
$\bar 8$-brane pairs.  There seems to be no gain in representing this 
D$p$-brane for example as a bound state of unstable 7-branes.  However, 
while not every configuration of stable D-branes of Type IIA fits into the 
worldvolume of a given 8-brane $\bar 8$-brane system, it clearly fits into 
the worldvolume of a system of spacetime-filling 9-branes.  This embedding 
will enable us to keep all spacetime symmetries manifest, and will not lead 
to any kinematical restrictions on configurations of lower-dimensional 
stable D$p$-branes that can be studied this way.  

Consider first a single unstable 9-brane in Type IIA theory.  The gauge group 
is $U(1)$, and the tachyon is just a real scalar field of charge zero.  We do 
not know the exact form of the tachyon potential $V(T)$, but we can assume 
that $V(T)=V(-T)$, and that $T$ will condense into one of two vacuum values, 
$T=\pm T_0$.  

We will assume -- in close analogy with a similar assumption made in \cite{sen} 
in the related case of pairs of stable $p$-branes and $\bar p$-branes -- that 
when the tachyon condenses into either $T_0$ or $-T_0$, the negative energy 
density associated with the condensate will cancel the positive energy density 
associated with the 9-brane tension, and the 9-brane will completely 
annihilate into the supersymmetric vacuum.  This 9-brane is an example of 
what we called an ``elementary'' 9-brane in the previous section.  

Since the vacuum manifold of the tachyon field consists of two points 
$\pm T_0$, the tachyon can form a kink of codimension one in spacetime.  Near 
the core of the kink, the tachyon field will be (up to a convergence factor, 
assuring that $T\rightarrow\pm T_0$ asymptotically at infinity) 
\begin{equation}\label{eetachkink}
T=\pm x^9,
\end{equation}
\noindent where $x^9$ is the coordinate normal to the core of the kink.  Thus, the 
core of this kink represents a domain wall in spacetime, which we will 
interpret as the supersymmetric 8-brane or $\bar 8$-brane, depending on the 
sign in \ref{eetachkink} (or, in other words, the sign of the difference between 
the asymptotic vacuum values of the tachyon $T(-\infty)-T(+\infty)$ on the two 
sides of the domain wall).  Notice that only one 8-brane or one $\bar 8$-brane 
can be constructed from one 9-brane.  

Consider now $N$ unstable 9-branes.  The gauge symmetry is $U(N)$, and the 
tachyon is in the adjoint of the gauge group.  The tachyon will again have 
the tendency to roll down to a certain value $T_0$ at a minimum of its 
potential, possibly breaking a part of the $U(N)$ gauge symmetry.  The 
precise pattern of symmetry breaking depends on the structure of eigenvalues 
of $T_0$, which in turn depends on the precise form of the tachyon potential, 
which of course is not known.  It is natural to expect that $V(T)=V(-T)$, and 
that all eigenvalues of $T_0$ are equal to a certain $T_v$, possibly up to a 
sign.  It is easy to see that this structure of eigenvalues would be obtained 
for example from even potentials of the form 
\begin{equation}\label{eenicepotent}
V(T)=-m^2 tr(T^2)+\lambda^2 tr(T^4)+\ldots,
\end{equation}
\noindent and with each term containing only a single trace.  Such terms in the 
potential are expected from the disc amplitudes, i.e.\ at tree level in 
open-string perturbation theory.  

When the tachyon on the worldvolume of $N$ 9-branes condenses into the vacuum 
value with $N-k$ positive eigenvalues and $k$ negative ones, 
\begin{equation}\label{eetachvacnk}
T_0=T_v\pmatrix{1_{N-k}&0\cr0&-1_k\cr},
\end{equation}
\noindent the $U(N)$ gauge symmetry is broken to $U(N-k)\times U(k)$.  Just as in the 
case of a single 9-brane, the tachyon field can form kinks of codimension 
one.  One particularly interesting case corresponds to the kink in all 
eigenvalues of $T$, localized at a common domain wall $Y$ of codimension one 
in spacetime, which near $Y$ can be written (again, up to a convergence 
factor), as 
\begin{equation}\label{eemultikink}
T=\pmatrix{x^9\cdot1_{N-k}&0\cr0&-x^9\cdot 1_k\cr}.
\end{equation}
\noindent We conjecture that this configuration should be interpreted as $N-k$ 8-branes 
and $k$ $\bar 8$-branes with coinciding wordvolumes wrapping $Y$.%
\footnote{The interpretation of this multiple kink configuration as a set of 
$8$-branes and $\bar 8$-branes suggests the existence of a coupling 
$$\int_XC_9\wedge tr(dT)$$
between the spacetime RR 9-form $C_9$ and the $U(N)$ tachyon $T$ on the 
9-brane worldvolume $X$.}
More general configurations of separated 8-branes and $\bar 8$-branes can 
be constructed by letting each eigenvalue vanish along a separate manifold of 
codimension one in spacetime.  

Thus, any number of 8-branes and $\bar8$-branes can be constructed from 
9-branes, but one cannot represent more than $N$ of them as a bound state of 
$N$ 9-branes: if we want to add an extra 8-brane, the construction has to be 
``stabilized'' in the sense of K-theory, by adding an extra 9-brane.  

In general, worldsheets with more than one boundaries could give rise to 
corrections to the tachyon potential \ref{eenicepotent} , of the form 
\begin{equation}\label{eenicecorrect}
\tilde\lambda^2(tr (T^2))^2+\ldots
\end{equation}
\noindent with more than one trace in each individual term.  Using the analysis of 
section~III.D. and Appendix B of \cite{lfli} , one can show that even in the 
case of a generic potential \ref{eenicepotent} and \ref{eenicecorrect} with 
$\lambda^2\geq 0$, $\tilde\lambda^2>0$, the minimum $T_0$ of the tachyon 
potential still has only two eigenvalues, $\pm T_0$, and the vacua with 
different values of $k$ stay degenerate.%
\footnote{I wish to thank John Preskill for bringing Ref.~\cite{lfli} to my attention.}

One could now combine the construction \ref{eemultikink} of 8-branes and 
$\bar 8$-branes from 9-branes with the construction discussed in 
\cite{ewk,sen}, and construct all lower-dimensional D-$2p$-branes as bound 
states of a sufficient number of 8-brane $\bar 8$-brane pairs prepared from 
9-branes.  

This two-step procedure has several significant drawbacks.  First of all, we 
have to select a preferred submanifold of codimension one in spacetime, which 
represents the worldvolume of the 8-brane $\bar 8$-brane system.  This breaks
some of the manifest spacetime symmetries.  Perhaps more importantly, 
configurations of lower-dimensional branes that cannot be embedded into the 
worldvolume of a single system of coincident 8-brane $\bar 8$-brane pairs may 
require -- when realized via the two-step construction involving 8-branes -- 
that extra 9-branes be introduced,  due to the fact that each 8-brane or 
$\bar 8$-brane needs its own 9-brane.  This would make the number of 9-branes 
used in the bound-state construction artificially dependent on the number 
and precise configuration of the lower-dimensional bound state.  

These shortcomings will be resolved in a one-step procedure, in which we 
construct arbitrary lower-dimensional D-$2p$-brane directly as a bound state 
of a system of 9-branes.  This one-step procedure avoids the intermediate 
step involving 8-branes and $\bar 8$-branes, and therefore avoids the 
degeneracy of the codimension-one bound-state construction, leading to a more 
powerful description of lower-dimensional D-branes as bound states.  Along 
the way, we will discover many intriguing connections to K-theory.  

\bigskip\noindent
{\it The General Bound State Construction}
\medskip

{}From now on, we will consider 9-brane systems whose tachyon condensate $T_0$ 
has an equal number of positive and negative eigenvalues.  Thus, the number 
of 9-branes is $2N$ for some $N$, and the gauge group $U(2N)$ is broken to 
$U(N)\times U(N)$.  The vacuum manifold is 
\begin{equation}\label{eevackone}
\CV_1(2N)=U(2N)/(U(N)\times U(N))
\end{equation}
\noindent We are interested in stable, vortex-like configurations in the tachyon field.  
Away from the core of such a stable vortex, the tachyon field (almost) assumes 
its vacuum values.  This defines a map of the sphere $\S^m$ surrounding the 
core of a vortex of codimension $m+1$ into the vacuum manifold $\CV_1(2N)$.  
Possible candidates for stable tachyon 
vortices in this codimension are thus classified by elements  of the 
homotopy group $\pi_m(\CV_1(2N))$.  It turns out that homotopy groups of the 
vacuum manifold \ref{eevackone} are non-trivial in even dimensions, 
$\pi_{2k}(\CV_1(2N))=\Z$, and trivial in all odd dimensions.  (Here we are 
assuming that $N$ is large enough, so that it belongs to the ``stable'' 
range.)  This should be contrasted with the case of Type IIB 9-brane 
$\bar 9$-brane pairs, reviewed in section~2.1:  in the case of Type IIB 
9-branes, the vacuum manifold $\CV_0(N)=U(N)$ has non-zero homotopy groups 
only in {\it odd\/} dimensions, $\pi_{2k-1}(\CV_0(N))=\Z$.  Therefore, while 
the Type IIB system of 9-brane $\bar 9$-brane pairs supports bound states of 
codimension $2k$, our Type IIA 9-brane system will exhibit bound states in 
codimensions $2k+1$. 

This structure of homotopy groups is not coincidental, and in fact reflects 
a deep connection of our construction to K-theory.  Our vacuum manifold 
$\CV_1(2N)$ can be thought of as a Grassmannian manifold whose points are 
$N$-dimensional complex subspaces in $\C^{2N}$.  This Grassmannian plays an 
important role in K-theory, as it represents a standard finite-dimensional 
approximation to the ``universal classifying space'' $BU$ (see e.g.\ 
\cite{husemoller}).  The importance of $BU$ in K-theory stems from the fact that 
the K-theory group $\tilde\K(X)$ is canonically isomorphic, for any 
(reasonable) $X$, to the set of homotopy classes of maps from $X$ to this 
universal classifying space, 
\begin{equation}\label{eeclassifk}
\tilde\K(X)=[X,BU].
\end{equation}
\noindent Similarly, the higher K-theory group $\K^{-1}(X)$ is related to the set of 
homotopy classes of maps from $X$ to the infinite unitary group $U$, 
\begin{equation}\label{eeclassifkone}
\K^{-1}(X)=[X,U].
\end{equation}
\noindent  Thus, the vacuum manifold $\CV_1(N)$ of the tachyon on Type IIA 9-branes 
is a finite-dimensional approximation to the classifying space $BU$, and the 
vacuum manifold $\CV_0(N)=U(N)$ of the tachyon in the Type IIB 9-brane 
$\bar 9$-brane system is a finite-dimensional approximation to the infinite 
unitary group $U$.  
   
The structure of homotopy groups of the tachyon vacuum manifold \newline $\CV_1(2N)$ 
indicates the possibility of bound states in all odd codimensions on the 
worldvolume of Type IIA 9-branes on $\R^{10}$.  These bound states will appear 
as tachyon vortices, and will be interpreted as supersymmetric D-$2p$-branes 
of Type IIA theory.  Just like in the case of Type IIB theory \cite{ewk} , K-theory 
suggests the number of Type IIA 9-branes that is particularly natural for the 
bound state construction.  Bound states of codimension $2k+1$ are most 
efficiently described by $2N=2^k$ 9-branes.  Stable tachyon vortices in this 
codimension are classified by the $2k$-th homotopy group of the vacuum 
manifold, $\pi_{2k}(\CV_1(2^k))=\Z$.  In fact, homotopy groups of $\CV_1$ are 
related to the homotopy groups of $U(N)$, via%
\footnote{This is precisely one half of the statement of Bott periodicity 
\cite{karoubi,atiyah,fomenko}.  The other half of Bott periodicity 
similarly relates odd homotopy groups of $\CV_1$ and even homotopy groups of 
$\CV_0$.  Using these relations, together with \ref{eeclassifk} and 
\ref{eeclassifkone} , one can for example derive all K-theory groups of spheres, 
used in section~2.2 to classify supersymmetric Type II D-branes in $\R^{10}$.}
\begin{eqnarray}\label{eebotthomot}
\ldots=\pi_{2k+1}(U(2^k))&=&\pi_{2k}(U(2^k)/U(2^{k-1})\times 
U(2^{k-1}))\nonumber \\ &=&\pi_{2k-1}(U(2^{k-1}))=\ldots.
\end{eqnarray}
\noindent  The tachyon vortex corresponding to the generator of $\pi_{2k}(\CV_1)$ can be 
explicitly constructed as follows.  The worldvolume of $2^k$ 9-branes supports 
a $U(2^k)$ Chan-Paton bundle, which we identify with the spinor bundle $\CS$ 
of the group $SO(2k+1)$ of rotations in the transverse dimensions.  The 
tachyon condensate is then given by the vortex configuration 
\begin{equation}\label{eetachkone}
T(x)=\Gamma_m x^m.
\end{equation}
\noindent  As in the Type IIB case \cite{ewk} , $\Gamma_m$ are the $\Gamma$-matrices of 
the group of rotations in transverse dimensions $x^m$, $m=1,\ldots, 2k+1$.  
\ref{eetachkone} describes a stable vortex in codimension $2k+1$, which we 
interpret as the supersymmetric $(8-2k)$-brane of Type IIA theory.   

Even though the expression for the tachyon vortex \ref{eetachkone} on 
Type IIA 9-branes looks formally identical to the tachyon vortex \ref{eetachk} 
on the system of 9-brane $\bar 9$-brane pairs of Type IIB theory, there is a 
significant difference between \ref{eetachkone} and \ref{eetachk} .   Asymptotically 
away from the vortex of the tachyon, \ref{eetachkone} takes values in 
the vacuum manifold $U(2^k)/(U(2^{k-1})\times U(2^{k-1}))$.  On the other 
hand, the Type IIB vortex \ref{eetachk} takes asymptotically values in 
$(U(2^{k-1})\times U(2^{k-1}))/U(2^{k-1})$.  One can see this distinction 
clearly in K-theory, where the two tachyon condensates represent generators 
of disctinct K-theory groups; \ref{eetachk} generates the relative group 
$\K(\B^{2k},\S^{2k-1})$ (with $\B^{2k}$ a ball $|x|^2\leq 1$ in $\R^{2k}$), 
and \ref{eetachkone} represents the generator of $\K^{-1}(\B^{2k+1},\S^{2k})$ 
\cite{karoubi,spingeo,abs}.  

The tachyon vortex \ref{eetachkone} is accompanied by a non-trivial $U(2^k)$ 
gauge field, due to the finite energy condition imposed on the whole 
configuration.  The non-triviality 
of the unbroken part of the gauge field is measured by the element in 
$\pi_{2k-1}(U(2^{k-1}))$ that maps to the element of $\pi_{2k}(\CV_1(2^k))$ 
corresponding to the tachyon condensate \ref{eetachkone} , under the isomorphism 
of homotopy groups \ref{eebotthomot} .  In the construction of the gauge field, 
one starts with topologically trivial gauge fields on the upper and lower 
hemisphere of $\S^{2k}$, and the element of $\pi_{2k-1}(U(2^{k-1}))$ 
corresponds to a large gauge transformation along the equator of $\S^{2k}$ 
that allows one to patch the gauge fields on the two coordinate systems into 
a gauge field carrying the corresponding topological charge on $\S^{2k}$.  
This is a standard construction known from the physics of magnetic monopoles 
\cite{thpolya} , and the unbroken long-range gauge field of $U(N)\times U(N)$ 
indeed corresponds to that of a generalized magnetic monopole.  The 
corresponding magnetic charge is measured by the $k$-th Chern class of the 
gauge bundle on $\S^{2k}$ (Some relevant background on magnetic monopoles can 
be found in \cite{monorevs,weinberg}.)  

The other side of the isomorphism \ref{eebotthomot} , which relates 
$\pi_{2k}(\CV_1(2^k))$ to $\pi_{2k+1}(U(2^k))$, is also important: it allows 
us to relate our bound-state construction to the definition of $\K^{-1}(X)$.  
Given the tachyon condensate $T=\Gamma\cdot x$, we can construct an element 
of $\pi_{2k+1}(U(2^k))$ as follows.  Consider
\begin{equation}\label{eetachbsauto}
\CU=-e^{\pi iT}.
\end{equation}
\noindent     

Since the tachyon is in the adjoint of $U(2^k)$, $\CU$ defines a map from the 
unit ball $|x|\leq 1$ to $U(2^k)$.  The group $U(2^k)$ as well as its Lie 
algebra can be represented by $2^k\times 2^k$ matrices which are unitary and 
hermitian, respectively.  We will use a particularly useful description of the 
coset \ref{eevackone} , as the set of all $2^k\times 2^k$ matrices that are 
simultaneously hermitian and unitary \cite{fomenko} .  In particular, all such 
matrices square to one, as elements in $U(2^k)$.  (Incidentally, this proves 
that far from the core of the vortex, for an appropriate convergence factor 
$f(|x|)$ omitted in \ref{eetachkone} , the tachyon condensate \ref{eetachkone} indeed takes values in the vacuum manifold $\CV_1(2^k)$.) 

We can now apply this understanding to the tachyon vortex \ref{eetachkone} .  
Using $T^2=|x|^2$, one can show that $\CU$ of \ref{eetachbsauto} maps the origin 
$x=0$ to $-1$ in $U(2^k)$, and each point with $x^2=1$ to the identity in 
$U(2^k)$.  Thus, \ref{eetachbsauto} indeed defines a map from 
$\S^{2k+1}$ to $U(2^k)$, and hence an element of $\pi_{2k+1}(U(2^k))$.  This 
element in $\pi_{2k+1}(U(2^k))$ maps under \ref{eebotthomot} to the element in 
$\pi_{2k}(\CV_1(2^k))$ that corresponds to the tachyon vortex \ref{eetachkone} .  
(For details on this K-theory construction, see \cite{fomenko} .)    

In terms of K-theory, this proves that our tachyon condensate actually 
represents the generator of the relative K-theory group 
$\K^{-1}(\B^{2k+1},\S^{2k})$, and our bound-state construction is precisely 
the analog of the ABS construction \cite{abs} , now mapping classes in $\K(Y)$ 
to classes in $\K^{-1}(X)$ for $Y$ of odd codimension in the spacetime 
manifold $X$ wrapped by the ustable 9-branes of Type IIA theory.%
\footnote{The relative K-theory group $\K^{-1}(\B^{2k+1},\S^{2k})$ is defined 
as the group of equivalence classes of bundles with automorphisms $(E,\alpha)$ 
on $\B^{2k+1}$, with $\alpha=1$ when restricted to the boundary $\S^{2k}$ 
(see e.g.\ \cite{karoubi} , section~II.3.25.).  In our string theory construction, 
$E$ is the Chan-Paton bundle, and as we have just seen, $\CU$ of 
\ref{eetachbsauto} has just the right properties to be identified with $\alpha$.}

This one-step construction of Type IIA D-branes as bound states of codimension 
$2k+1$ in a system of unstable 9-branes suggests the following hierarchy of 
bound state constructions.  Consider a supersymmetric D$p$-brane in Type IIA 
or Type IIB theory.  This D-brane can be constructed as a bound state (tachyon 
kink) in the worldvolume of an unstable D-$(p+1)$-brane.  Alternatively, 
it can be constructed \cite{sen,ewk} as a bound state of a $(p+2)$-brane 
$\bar{(p+2)}$-brane pair.  It can also be constructed as a bound state of two 
unstable $(p+3)$-branes, etc.  This hierarchy of brane systems of increasing 
dimensions supports worldvolume gauge groups that form a natural hierarchy, 
\begin{equation}\label{eegaugehier}
U(1)\ \subset\ U(1)\times U(1)\ \subset\ U(2)\ \subset\ 
U(2)\times U(2)\ \subset\ U(4)\ \subset\ U(4)\times U(4)\ \ldots
\end{equation}
\noindent     

In this hierarchy, the bound state construction of \cite{sen,ewk} in terms 
of pairs of stable branes alternates with the bound state construction in 
terms of ustable branes presented in this section.  This procedure can be 
iterated until we reach the spacetime-filling dimension, thus ending up with 
a construction in terms of 9-branes where all spacetime symmetries are 
manifest.  

\bigskip\noindent
{\it Codimension Three: The 't~Hooft-Polyakov Monopole}
\medskip

It is instructive to look more closely at the construction of bound states 
of codimension $2k+1=3$.  The gauge group suggested by K-theory is $U(2)$, 
acting on a pair of unstable branes whose Chan-Paton bundle in {\bf 2} of 
$U(2)$ is identified with the two-dimensional spinor bundle $\CS$ of the 
$SO(3)$ group of rotations in the transverse spacetime dimensions.  
The tachyon is a map from the two-dimensional spinor bundle $\CS$ back to 
$\CS$.  Using the $\Gamma$-matrices $\sigma_i$ of $SO(3)$, the vortex of 
vorticity one can be written (up to an overall normalization factor) as 

\begin{equation}\label{eevortextd}
T=\pmatrix{x^3&x^1+ix^2\cr x^1-ix^2&-x^3\cr}.
\end{equation}
\noindent    The finite-energy condition ties \ref{eevortextd} to a non-trivial gauge field, 
which takes the form
\begin{equation}\label{eegaugetd}
A_i=f(|x|)\Gamma_{ij}x^j,
\end{equation}
\noindent   where $f(|x|)$ is a known convergence factor \cite{thpolya} , and $\Gamma_{ij}$ is 
the antisymmetrized product of the $SO(3)$ $\Gamma$-matrices $\sigma_i$ and 
$\sigma_j$.  Up to the trivial lift from $SU(2)$ to $U(2)$ gauge theory, this 
is precisely the 't~Hooft-Polyakov magnetic monopole in $3+1$ dimensions 
\cite{thpolya} !  In our construction, this monopole represents the supersymmetric 
stable D-$2p$-brane of Type IIA theory as a bound state of two unstable 
D-$(2p+3)$-branes.  

On the off-diagonal in \ref{eevortextd} we recognize the tachyon condensate 
\ref{eetachcodimtwo} that appeared in the construction \cite{sen,ewk} of the 
D-$2p$-brane as a bound state of a $(2p+2)$-brane $\bar{(2p+2)}$-brane pair.  
Similarly, on the diagonal in \ref{eevortextd} we find the tachyon kink 
\ref{eemultikink} that corresponds to the construction of one $(2p+2)$-brane and 
one $\bar{(2p+2)}$-brane from a pair of $(2p+3)$-branes.  Thus, we can see 
that our one-step bound state construction can also be interpreted as a 
two-step construction, whereby we first prepare a $(2p+2)$-brane 
$\bar{(2p+2)}$-brane pair, which then forms a $2p$-brane bound state.  
In this two-step construction, however, we lose some of the manifest 
symmetries of \ref{eevortextd} by choosing an embedding of the $2p$-brane 
worldvolume into a worldvolume of the $(2p+2)$-brane $\bar{(2p+2)}$-brane 
pair.  
  
\bigskip\noindent
{\it Comments on Codimension One}
\medskip

In codimension one, i.e.\ for $k=0$, the suggested value $2^k$ of the number 
of unstable Type IIA branes is one.  Thus, there is no symmetry breaking of 
the $U(1)$ gauge symmetry, and we are left with a $U(1)$ gauge theory and a 
tachyon that can condense into either one of the two vacuum values, 
$\pm T_0$.  This case was discussed at the beginning of this section, in the 
case of 8-branes and $\bar 8$-branes as tachyon kinks on 9-brane 
worldvolumes.  The symmetry restored at the core of the kink is the $\Z_2$ 
symmetry $T\rightarrow -T$.  

We can now see from the K-theory perspective why this case of codimen-sion-one
bound states is rather degenerate.  Indeed, the relation between homotopy 
groups of the vacuum manifold and those of the unitary groups, as given for 
$k\geq 1$ in \ref{eebotthomot} , becomes degenerate for $k=0$.  The relevant 
homotopy group of the vacuum manifold for one 9-brane is $\pi_0(\pm T_0)=
\Z_2$, and there is not enough room for the anticipated conserved 8-brane 
charge that should be classified by $\Z$.  Thus, each individual 8-brane 
or $\bar 8$-brane requires an extra 9-brane; the smallest 9-brane system 
that would accomodate the full $\K^{-1}(\S^1)=\Z$ group of 8-brane charges 
would contain an infinite number of 9-branes.  

This also sheds some light on the two-step construction of the bound 
state.  Consider a bound state of codimension $2k+1$, as described in 
\ref{eetachkone} .  Choose $Y$ of codimension one, such that the worldvolume of our 
bound state lies in $Y$.  Split the transverse dimensions as $2k$ plus one, 
where the $2k$ dimensions are within $Y$, and the last dimension is normal to 
$Y$.  The tachyon vortex \ref{eetachkone} can be written as 
\begin{equation}\label{eetachsplit}
T=\pmatrix{x^9&\Gamma\cdot x\cr\Gamma^\dagger\cdot x&-x^9},
\end{equation}
\noindent where $x^9$ parametrizes the dimension transverse to $Y$, and $\Gamma$ are 
the $\Gamma$-matrices of $SO(2k)$ along $Y$.  \ref{eetachsplit} has a nice 
intuitive interpretation: the terms on the diagonal look 
precisely like $2^{k-1}$ 8-branes and $2^{k-1}$ $\bar 8$-branes, and the 
off-diagonal terms correspond precisely to the bound state in the sense of 
\cite{ewk} , which represents our brane as a bound state of codimension $2k$ in 
a system of $2^{k-1}$ 8-brane $\bar 8$-brane pairs.  

However, it is clear that such a two-step construction is artificial.  Not 
only does it break manifest $SO(2k+1)$ rotation invariance in the $2k+1$ 
dimensions transverse to the worldvolume of our brane; it also artificially 
relates the non-degenerate case of $k\geq 1$ to the degenerate case of bound 
states of codimension one.  

\bigskip\noindent
{\it Comparison to Bound States of Codimension One in Type I Theory}
\medskip

Superficially, the construction presented above appears somewhat reminiscent of
a construction \cite{sen,ewk} in orientifolds of Type IIB theory, where 
certain stable non-supersymmetric D$p$-branes carrying $\Z_2$ charges are 
realized as bound states of a number of pairs of stable $(p+1)$-branes and 
$\bar{(p+1)}$-branes.  (Sen considers $p=0$ \cite{sen} , while $p=8$ is discussed 
in \cite{ewk} .)  The following remark is intended to clarify the distinction 
between the two constructions.  

We have seen that the tachyon condensate \ref{eetachkone} on the worldvolume of 
unstable 9-branes of Type IIA theory represents a generator of the relative 
K-theory group $\K^{-1}(\B^{2k+1},\S^{2k})=\Z$, and the bound state 
construction of a D-brane wrapping a submanifold $Y$ of codimension $2k+1$ in 
spacetime $X$ represents a map from $\tilde\K(Y)$ to $\K^{-1}(X)$.  More 
precisely, this embedding of a lower-dimensional brane into the 9-brane system 
corresponds to the isomorphism 
\begin{equation}\label{eethomkone}
\tilde\K(Y)\otimes\K^{-1}(\B^{2k+1},\S^{2k})\cong
\K^{-1}(Y\times\B^{2k+1},Y\times\S^{2k}),
\end{equation}
\noindent where $Y\times\B^{2k+1}$ is a small neighborhood of $Y$ in spacetime.  
The tachyon \ref{eetachkone} is a convenient representation of the generator of 
$\K^{-1}(\B^{2k+1},\S^{2k})$, and the isomorphism between $\tilde\K(Y)$ and 
$\K^{-1}(Y\times\B^{2k+1},Y\times\S^{2k})$ is realized by the cup product with 
this generator.  

In contrast, the tachyon kink used in \cite{sen} to describe the stable 
$\Z_2$-charged D0-brane of Type I theory as a bound state of a 1-brane 
$\bar 1$-brane pair, represents the generator of $\KO(\B^1,\S^0)=\Z_2$.  The 
cup product with this generator maps $\Z_2$ classes in KO-theory on $Y$ 
(representing the D0-brane carrying a $\Z_2$ charge) to $\Z_2$ classes in 
$\KO(Y\times\B^1,Y\times\S^0)$.

\subsection{Type $I'$ Theory and $\KR^{-1}(X)$}
One can generalize the construction to the Type $\I'$ orientifold of Type IIA 
theory.  The generalization is relatively straightforward, and we will be 
very brief.  

Consider the orientifold of Type IIA theory on $\R^{10}$, with the orientifold 
group $\Z_2$ that changes the orientation of one spacetime dimension.  
This theory contains unstable spacetime-filling 9-branes, whose configurations 
up to creation and annihilation of elementary 9-branes classify all possible 
D-brane charges.  In terms of K-theory, this corresponds to a group called 
$\KR^{-1}(X)$ \cite{mfareality,spingeo,ewk}, which can be defined as the 
group of equivalence classes of pairs $(E,\alpha)$, where $E$ is a bundle with 
an antilinear involution that commutes with the orientifold group, and 
$\alpha$ is an 
automorphism on $E$ that also preserves the orientifold group action.  In 
terms of Type $\I'$ string theory, $E$ again corresponds to the Chan-Paton 
bundle on the worldvolume of the spacetime-filling 9-branes.  At the 
orientifold planes, the gauge group is reduced from the unitary group to its 
orthogonal subgroup.  

Each individual lower-dimensional D-brane can be represented as a bound state 
of a certain number of Type $\I'$ 9-branes.  The tachyon condensate is 
required to respect the orientifold $\Z_2$ symmetry, and therefore corresponds 
to what might be called a ``$\Z_2$-equivariant monopole.'' (Similar 
equivariant solitons and instantons were studied in \cite{etsm} .)  

Far away from the orientifold planes, we can think of the spacetime manifold 
$X$ as being represented by a double cover $\tilde X$ of $X$ with the 
orientifold group mapping the two disconnected components of $\tilde X$ to 
each other.  Using a standard result of K-theory \cite{mfareality,spingeo}
\begin{equation}\label{eekrk}
\KR^{-1}(\tilde X)\cong \K^{-1}(X),
\end{equation}
\noindent  we can see that far away from the orientifold planes, we recover the Type IIA 
construction that occupied most of this section.   

\section{K-Theory and Matrix Theory}
\setcounter{equation}{0}

Consider $N$ D0-branes in Type IIA theory on $\R^{10}$, or a toroidal 
compactification thereof.  (This restriction is mostly for simplicity; the 
construction can be straightforwardly extended to more complicated 
compactifications as well.)  Using the general bound-state construction of 
section~3, each D0-brane can be described as a bound state of sixteen unstable 
spacetime-filling 9-branes.  

The worldvolume field theory of sixteen 9-branes contains a $U(16)$ gauge 
field, a tachyon $T$ in the adjoint of $U(16)$, and a pair of chiral fermions 
$\chi$ and $\chi'$ of opposite chiralities, also in the adjoint of $U(16)$.  
In addition, we have the usual hierarchy of massive string states, and this 
system will couple to the Type IIA closed string sector.  

We start with a configuration of sixteen 9-branes, in which the tachyon rolls 
down to a minimum $T_0$, with eight positive and eight negative eigenvalues 
$\pm T_v$.  This condensate breaks $U(16)$ to $U(8)\times U(8)$, and the 
vacuum manifold is $\CV=U(16)/(U(8)\times U(8))$.  Since $\pi_8(\CV)=\Z$, 
the tachyon can develop a stable point-like vortex, whose form near the core 
can be described by 
\begin{equation}\label{eetachdzero}
T=\Gamma_mx^m.
\end{equation}
\noindent    Here, again, $\Gamma_m$ are the $\Gamma$-matrices of $SO(9)$, the group of 
rotations in the dimensions transverse to the core of the vortex.  This 
configuration carries vorticity one.  In order to keep the energy of this 
localized object finite, the long-range gauge field will also be non-trivial, 
and will in fact give rise to a non-zero ``magnetic charge'' of the object.  
(More precisely, the corresponding 4-th Chern class as measured by the gauge 
field on the 8-sphere surrounding the core of the vortex will be equal to 
one.)  

The sixteen 9-branes are in the {\bf 16} of the gauge group $U(16)$.  In the 
background of the generalized magnetic monopole/vortex representing the 
D0-brane, this {\bf 16} is identified with the spinor {\bf 16} of the $SO(9)$ 
spacetime rotation symmetry in the dimensions transverse to the worldvolume 
of the D0-brane.  This represents a higher-dimensional generalization of the 
well-known phenomenon in three space dimensions, where the background of the 
$SU(2)$ 't~Hooft-Polyakov monopole identifies the {\bf 3} of the gauge group 
with the {\bf 3} of the space rotation group $SO(3)$.  

A scaling argument \cite{weinberg} clearly suggests that the configuration will 
lower its energy by shrinking its core, and our description in terms of 
low-energy field theory on the 9-brane worldvolume will cease to be 
adequate.  However, purely on topological grounds, one will be left with 
a stable soliton in string theory.  The K-theory origin of this construction 
indicates that the soliton carries one unit of D0-brane charge, and therefore 
represents a D0-brane as a bound state of sixteen unstable 9-branes.  

So far we have found the bound state that describes one D0-brane.  Imagine 
now that we are interested in a system of $N$ D0-branes.  There is no reason 
why we should use a new set of sixteen 9-branes for each individual D0-brane.  
(In terms of K-theory, the trivial topology of the D0-brane worldlines in 
flat spacetime does not require ``stabilization'' of the configuration by 
adding extra 9-branes.)  Thus, in order to describe $N$ D0-branes, {\it we do 
not have to add sixteen extra 9-branes each time we add a D0-brane -- they 
can all be represented as bound states in a fixed system of sixteen 
spacetime-filling 9-branes.}

Since we can construct a multi-D0-brane state using just sixteen Type IIA 
9-branes, we can follow this 9-brane configuration as we take the scaling 
limit \cite{bfss},\cite{dkps}-\cite{mmns} that defines Matrix theory.  This suggests the 
intriguing possibility that we can formulate Matrix theory as a theory of 
stable solitons on the space-time filling worldvolume of sixteen unstable 
9-branes!   We can also add higher-dimensional D-branes to the system of D0-branes on 
$\R^{10}$, as vortices of codimension $2k+1$ in the 9-brane worldvolume.  
These vortices are stable, since 
\begin{equation}\label{eehomotmatrix}
\pi_{2k}(U(16)/(U(8)\times U(8)))=\Z
\end{equation}
(while all odd homotopy groups vanish).  They are naturally formed as bound 
states of a number of 9-branes that is typically smaller than sixteen 
(in fact, the results of section~3 suggest that the number of 9-branes 
involved in a bound state of codimension $2k+1$ is $2^k$), while the rest of 
the sixteen 9-branes are spectators in this construction.  We can also use 
the bound state construction to study the full system of stable branes of 
various dimensionalities on a general spacetime manifold $X$.  However, due to 
the non-trivial topology, this general case may require that extra 9-branes be 
added to the sixteen 9-branes that we have used in the description of 
D0-branes.  The general construction would require the full K-theory 
construction as discussed in the case of Type IIB theory in \cite{ewk} .  The 
``stabilization'' by addition of 9-branes can be avoided for the system of 
$N$ D0-branes in flat spacetime, due to the trivial topology of the D0-brane 
worldlines.  

This possible reinterpretation of Matrix theory in terms of vortices in a 
gauge theory with fixed gauge group is intriguing, since it allows us to 
change the number od D0-branes in the system without changing the rank of the 
gauge group.  In Matrix theory, we would like to understand how systems with 
different values of $N$ are related to each other, possibly via some RG-like 
relation.  This problem is notoriously difficult in the conventional 
formulation of Matrix theory, as it requires relating theories with gauge 
groups of different ranks.  In contrast, the K-theory construction of 
D0-branes as magnetic vortices keeps the gauge group fixed for arbitrary 
values of $N$.  

The non-supersymmetric $U(16)$ gauge theory on the worldvolume of sixteen 
9-branes is defined through its embedding into the supersymmetric Type IIA 
string theory.  In particular, it is unclear whether there are any useful 
limits in which supergravity decouples and leaves behind a system defined 
purely in terms of 9-brane degrees of freedom.  Nevertheless, one might 
expect that at least for compactifications down to four dimensions or lower, 
interesting decoupling limits might exist, whereby D0-branes appear as 
magnetic vortices in a decoupled non-supersymmetric gauge theory.  

It would be interesting to see directly in the bound-state construction how 
the system of $N$ D0-branes effectively decompactifies the eleventh dimension 
of Matrix theory \cite{bbpt,bgl}, and forms a ``bubble'' of the 
eleven-dimensional spacetime.  Recall from section~2.2 that $\K^{-1}(X)$ is 
usually defined by starting with the K-theory of $X\times \S^1$, and imposing 
an extra condition that restricts the Type IIA D-brane charges to a subgroup 
$\K^{-1}(X)\subset\tilde\K(X\times\S^1)$.  In other words, one can define 
$\K^{-1}(X)$ as K-theory of an eleven-dimensional extension of $X$, which is 
however not $X\times\S^1$ but rather $S'(X)$, defined as $X\times I$ with 
each boundary component ``pinched'' into a point.  The replacement of 
$X\times\S^1$ by $S'(X)$ is necessary to eliminate classes 
in $\tilde\K(X\times\S^1)$ that are not in $\K^{-1}(X)$.  
{}For example, the one-point compactification of the ten-dimensional 
spacetime $\R^{10}$ is $\S^{10}$, but its suspension $S'(\R^{10})$ is an 
eleven-sphere.  In combination with Matrix theory, the string theory 
construction of $\K^{-1}(X)$ in terms of bound states in 9-branes may perhaps 
alleviate some of the mystery related to the ``M-theory'' definition of 
$\K^{-1}(X)$ in terms of $\tilde\K(X\times\S^1)$.  

The construction of multi-D0-brane systems from a fixed system of 9-branes, 
relevant to Matrix theory, has an analog in the case of D-strings of Type I 
theory \cite{ewk} .  It was pointed out in \cite{ewk} that the Fock space of perturbative 
heterotic string theory should be contained in the system of eight 9-brane 
$\bar 9$-brane pairs.  Thus, the construction of \cite{ewk} should have similar 
implications in heterotic Matrix string theory \cite{hetmatrix} .  

\bigskip\noindent
{\it Comparison to Holographic Field Theory}
\medskip

There is a number of intriguing similarities between this K-theory-inspired 
construction of a system of D0-branes in terms of vortices in a $U(16)$ gauge 
theory on the worldvolume of unstable 9-branes of Type IIA theory, and the 
ideas of holographic field theory suggested in the context of non-perturbative 
M-theory in \cite{hft} .  Whether they are indicative of some closer relation 
remains to be seen.  

Here is a list of some of them:

(1) In K-theory, D0-branes appear as vortices (or generalized magnetic 
monopoles) in a gauge theory with fixed rank.  Similarly, in holographic field 
theory, the partons (to be compared to D0-branes of Matrix theory) appear 
as vortices in a fixed-rank gauge group.  Thus, in both cases, the limit to 
be compared to matrix theory requires one to look at multi-particle systems 
of vortices in a fixed-rank gauge group.  D0-branes would appear as solitonic 
excitations in a medium not dissimilar to some condensed matter systems.  

(2) The gauge group of holographic field theory is 
\begin{equation}\label{eehftgg}
\CG=\OSp(1|32)\times\OSp(1|32).
\end{equation}
\noindent  For many topological purposes, this group can be considered equivalent to 
its maximal compact subgroup.  The maximal compact subgroup of this 
non-compact version of $\OSp(1|32)$ is, in fact, $U(16)$.  When extended 
beyond ten dimensions, the natural hierarchy of gauge groups \ref{eegaugehier} 
suggests $U(16)\times U(16)$ as the group relevant to eleven dimensions.  
$U(16)\times U(16)$ is the maximal compact subgroup of the gauge group 
\ref{eehftgg} of holographic field theory.  

(3) The gauge group $\CG$ can be interpreted as the minimal extension of 
AdS group in eleven dimensions compatible with supersymmetry and parity 
invariance.  Perhaps more interestingly, it can also be viewed as a non-chiral 
Lorentz group in $(10,2)$ dimensions \cite{hft} .  This leads to a convenient 
representation of the bosonic subgroup $\Sp(32,\R)\times\Sp(32,\R)$ in terms 
of $64\times 64$ $\Gamma$-matrices acting on the spinor bundle 
$\CS=\CS_+\oplus\CS_-$ whose sections are 64-component non-chiral spinors 
in $(10,2)$ dimensions.  Each $\Sp(32,\R)$ acts on one of the two factors in 
$\CS_+\oplus\CS_-$.  This is again very reminiscent of structures appearing 
in K-theory.  

(4) The previous point can be extended even further.  The construction that 
embeds lower-dimensional branes as bound states into systems of 
higher-dimensional branes is of course a known K-theory construction \cite{abs} .  
The precise mathematical form of this map is realized via the Euler class 
of the normal bundle of the embedding of the lower-dimensional brane into 
the worldvolume of the higher-dimensional branes (see, e.g., \cite{karoubi} , 
sect.~IV.1).  The Lagrangian of holographic field theory \cite{hft} is intimately 
related to the Euler class.  More exactly, the Lagrangian can be interpreted 
as the Euler class of the tangent space to the twelve-dimensional manifold 
that has the eleven-dimensional spacetime as its boundary: the exterior 
derivative of the Chern-Simons Lagrangian of \cite{hft} is a supersymmetrization 
of the Euler density in $(10,2)$ dimensions.

\noindent It is a pleasure to thank Oren Bergman, Eric Gimon, Djordje Minic, Michael 
Peskin, John Preskill, John Schwarz, Steve Shenker, Eva Silverstein, 
Lenny Susskind and Edward Witten for valuable discussions.  I wish to thank 
the Stanford Institute of Theoretical Physics for hospitality during some 
parts of this work.  This work has been supported by Sherman Fairchild Prize 
Fellowship and by DOE Grant DE-FG03-92-ER~40701.

\end{document}